\documentclass[11pt,twoside]{article}
\usepackage{asp2004}
\usepackage{psfig}
\usepackage{epsf}
\usepackage{graphics}
\usepackage{lscape}
\def\edcomment#1{\iffalse\marginpar{\raggedright\sl#1\/}\else\relax\fi}
\reversemarginpar

\begin{document}
\thispagestyle{plain}

\title{Infrared Properties of Cataclysmic Variables in the 2MASS All 
Sky Data Release}
\author{D.\ W. Hoard$^1$, C.\ S.\ Brinkworth$^{1,2}$, S.\ Wachter$^1$}
\affil{$^1$Spitzer Science Center, California Institute of Technology, 
Pasadena, CA 91125, USA\\
$^2$Department of Physics and Astronomy, University of Southampton, 
Highfield, Southampton SO17 1BJ, UK}

\begin{abstract}
We present results from our analysis of the near-infrared ($J$, $H$, 
and $K_s$) photometry for all cataclysmic variables (CVs) from the 
catalog of \citet{D01} that are detected in the (final) All Sky Data 
Release (ASDR) from the Two Micron All Sky Survey (2MASS).  
\end{abstract}

\section{Introduction}

CVs have been ``traditionally'' observed primarily at short wavelengths 
because accretion-generated luminosity, which peaks in the 
optical--ultraviolet, dominates the radiated energy of most systems.  
Hence, relatively little is known about their infrared (IR) properties.  
Investigating CVs in the IR contributes to the understanding of system 
components that are expected to radiate at these wavelengths, such as 
the cool outer disk, accretion stream, and secondary star.   
In \citet{hoard02}, we presented an initial study of the group IR 
properties of CVs using the 2MASS 2nd Incremental Data Release (2IDR), 
drawn from 525 valid targets located in the 2IDR sky coverage.  We 
present here the preliminary results obtained from extending this study 
to the 2MASS ASDR.

\section{Data Properties}

Our input target list for the 2MASS ASDR consisted of the 1320 valid CVs 
listed in \citet{D01} as of 02 February 2004.  Each CV was securely 
identified using optical finding charts from \citet{D01} and/or other 
literature sources before matching it to an object in the 2MASS ASDR 
images and point source catalog.  We could not recover the CV for 239 
targets (typically CVs in crowded star clusters, old novae near the 
Galactic plane, etc.) and could not securely identify the IR counterpart 
for 104 (typically in regions where the IR field is very crowded).  The 
locations of another 376 CVs were securely identified, but these systems 
were too faint to be detected by 2MASS.  The remaining 601 CVs were 
detected by 2MASS; of these, we classify 362 as ``good'' detections 
($1\sigma$ uncertainties less than 0.1 mag for $J$, $H$, and $K_s$), 
165 as ``moderate'' detections ($1\sigma$ uncertainty larger than 0.1 mag 
for at least one of $J$, $H$, or $K_s$), and 74 as ``poor'' detections 
(lacking a formal uncertainty for at least one of $J$, $H$, or $K_s$, 
indicating that the target is near the 2MASS faint detection limit).

\section{Results}

\begin{figure}[tb!]
\plottwo{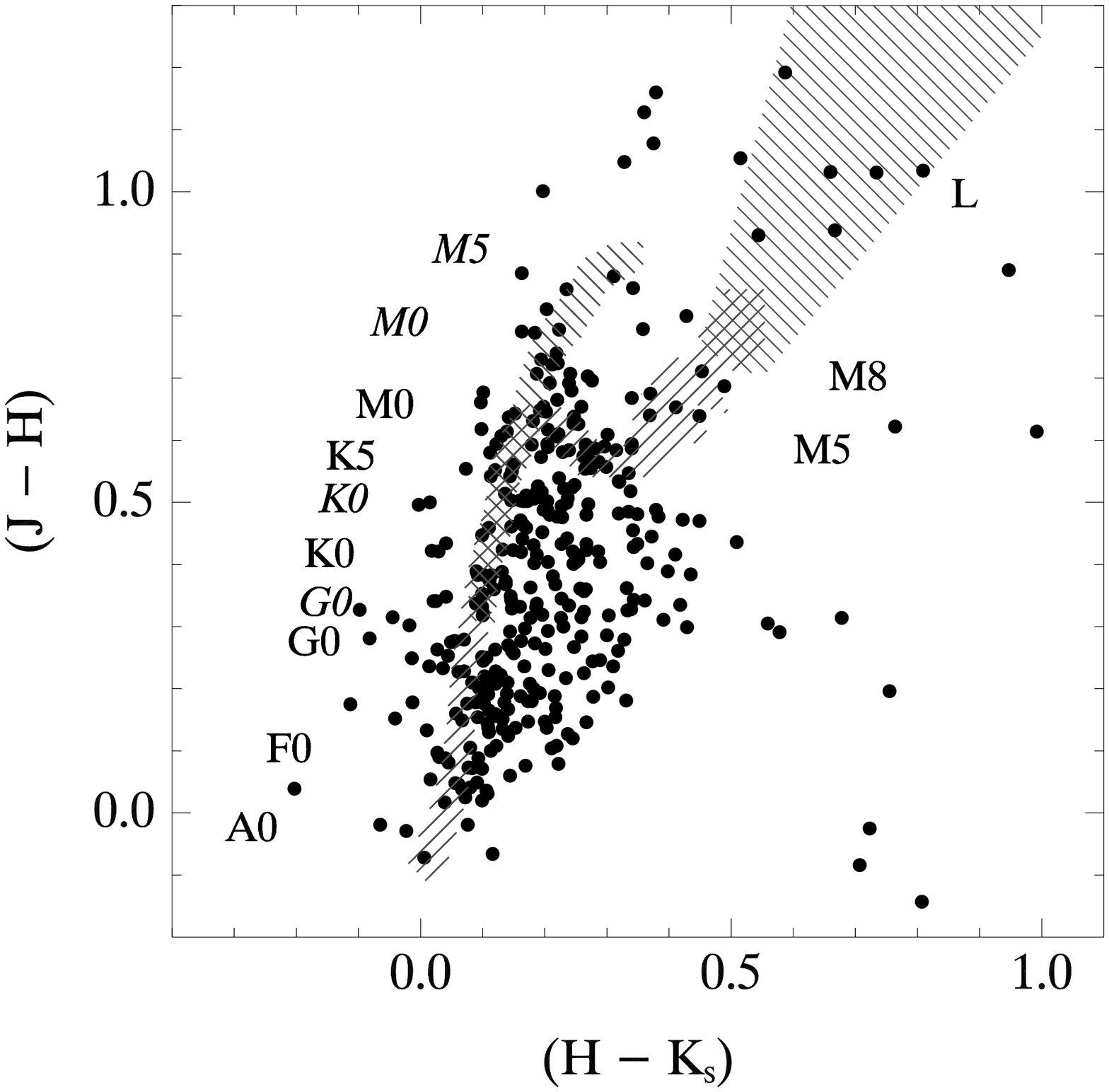}{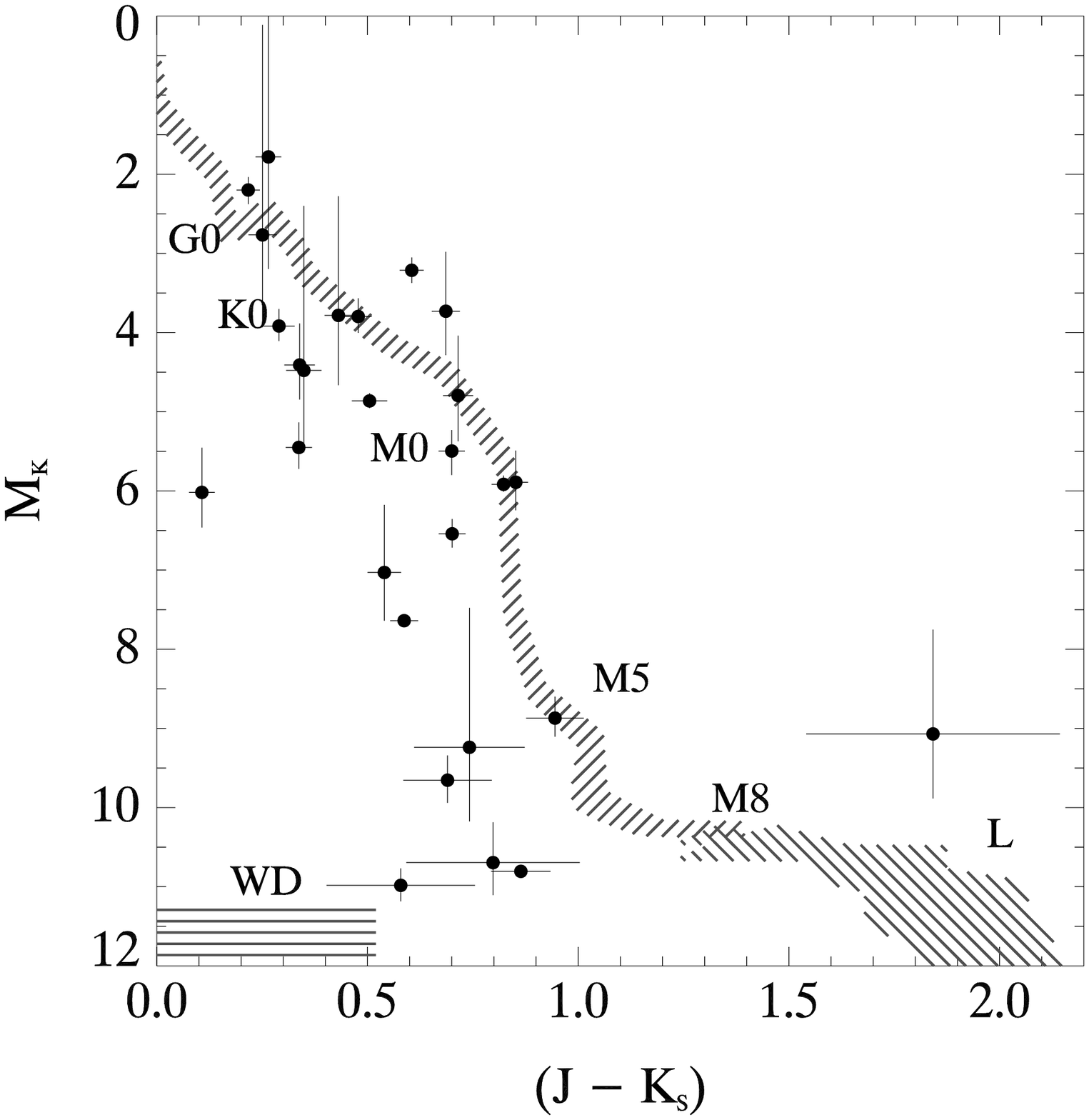}
\caption{(left) IR color-color diagram for all CVs classified as ``good'' 
detections (see text) in 2MASS.  (right) IR color-magnitude diagram for 
CVs detected in 2MASS, with distances determined from trigonometric 
parallax.}
\label{fig1}
\end{figure}

Figure \ref{fig1} (left) shows the IR color-color diagram of all ``good'' 
CV detections from 2MASS.  The loci of the main sequence (MS), L dwarfs, 
and giant stars are shown as cross-hatched regions labelled with spectral 
types (offset horizontally) in normal (MS, L dwarfs) and italic (giants) 
fonts.  The CVs located in outlying color regions (e.g., L dwarf region, 
lower right quadrant, etc.) may be mis-identified non-CVs or potentially 
interesting systems.  Figure \ref{fig1} (right) shows the IR 
color-magnitude diagram for CVs with distances determined from 
trigonometric parallax (e.g., \citealt{harrison99,thor03}; etc.).  The 
object closest to the L dwarf region is EF Eri, a CV that is believed 
to contain a brown-dwarf-like secondary star \citep{howell01}.
We are currently updating the input target sample to include all CVs 
in \citet{D01} up to 01 September 2004 for final analysis of the IR 
properties of CVs, to appear in a future publication.

\acknowledgments{This research was carried out at the Jet Propulsion 
Laboratory, California Institute of Technology, and is based on work 
supported by the National Aeronautics and Space Administration under 
an ADP grant issued through the Office of Space Science.  CSB acknowledges 
support from the Spitzer Science Center Visiting Graduate Student Program.  
This work uses data products from 2MASS, a joint project of the U.\ of 
Massachusetts and IPAC/Caltech, funded by NASA and the NSF.}

\end{document}